%% file: main.tex
\documentclass[letterpaper]{article}
\usepackage[preprint]{aaai2027}
\usepackage[hyphens]{url}
\usepackage{graphicx}
\usepackage{natbib}
\usepackage{caption}
\usepackage{booktabs}
\usepackage{tabularx}
\usepackage{xcolor}
\usepackage{colortbl}
\newcommand{\stack}[1]{\begin{tabular}[c]{@{}c@{}}#1\end{tabular}}

\usepackage[colorinlistoftodos]{todonotes}
\usepackage{pifont}
\definecolor{figagent}{HTML}{6B4FA0}
\definecolor{figharn}{HTML}{8A5A1E}
\definecolor{figtl}{HTML}{3D8B5A}
\definecolor{figevid}{HTML}{2F6FB0}
\newcommand{\stepref}[2]{\textcolor{#1}{\ding{\numexpr181+#2\relax}}}

\newcommand{\appendixsection}[2]{%
    \refstepcounter{section}%
    \section*{Appendix \thesection: #1}\label{#2}%
}
\graphicspath{{figures/}}

\title{Cassette-Editor: Shared Timeline Semantics for Human and Agent Editing}
\author{
    Zhixin Zeng\textsuperscript{\rm 1},
    Jiahao Zhong\textsuperscript{\rm 2}
}
\affiliations{
    \textsuperscript{\rm 1}Independent Researcher\\
    \textsuperscript{\rm 2}Nanyang Technological University, Singapore\\
    zeng.zhix@northeastern.edu, jzhong017@e.ntu.edu.sg
}

\begin{document}
\maketitle

\begin{abstract}
\input{sections/abstract}
\end{abstract}

\input{sections/introduction}
\input{sections/related_work}
\input{sections/background}

\input{sections/system}
\input{sections/evaluation}
\input{sections/limitations}

\input{sections/conclusion}

\clearpage
\bibliography{references}

\appendix
\input{sections/appendix}

\end{document}

%% file: sections/abstract.tex
Video editing involves tedious manual manipulation across interconnected timeline tracks, motivating automated agents that offload routine work while keeping the project directly revisable by humans. Fixed operation catalogs limit expressiveness, while even a basic trim requires coordinated updates across later clips, attached titles, and transitions. We present \textbf{Cassette-Editor}, an agent harness that unifies manual user gestures and automated agent patches within a shared editing environment. The shared action language expresses the editor's manual content changes, and a common history records committed edits for undo and manual refinement. When the model selects an edit, a state executor automatically derives its downstream consequences, such as shifting later clips and attached titles after a trim, while a validation rulebook enforces structural invariants and returns localized diagnostics for automated repair. To keep the agent grounded as the project evolves, dynamic read operations map transcripts and shot descriptions onto the current timeline. Evaluated with GPT-6 Luna, the full harness passed the scored request checks in 92/93 trials across 31 authored requests on two interview documentaries, compared with 48/93 for an internal basic-tools baseline. In ablations, disabling the consequences derived by the executor reduced passed trials from 230/232 to 199/232 on edits with dependent timeline elements, while ablating rulebook validation and repair lowered passed trials from 46/48 to 10/48 on requests operating near structural validity boundaries. Evaluating the complete system on 50 TVSum videos further demonstrated greater agreement with human annotations than a general agent using the same model (F1 23.8 vs. 20.1).

%% file: sections/introduction.tex
\section{Introduction}
\input{figures/fig-aci-overview}

Timeline video editing is an iterative process in which every cut, title, and transition remains open to inspection, revision, and undo. Because timeline tracks are tightly coupled, even a basic change such as shortening a storyline clip can shift later clips, displace attached titles, and alter the source frames available to a transition. Bringing a language agent into this workflow therefore requires dividing responsibility between the model and the editor when manual gestures and agent actions modify the same persistent project.

Existing LLM-based video editors invoke catalogs of editing operations \citep{wang2024lave,tilekbay2024expressedit,sandovalcastaneda2025editduet}, generate plans or scripts that a pipeline renders into media \citep{ding2025promptdriven,nath2026unified,zhao2026cutclaw,yan2026crayotter}, or act through an editor's graphical interface or through code \citep{hu2026cutverse,gupta2026timelinebench}. LAVE and ExpressEdit already combine agent actions with manual editing. We focus on how the editor maintains shared execution semantics and observations as both writers modify the project, with three design requirements.

First, the agent needs \emph{expressive parity over content changes}: its language must be able to express the same project changes a user can make through manual editing. Participants in ExpressEdit felt constrained by its seven operations \citep{tilekbay2024expressedit}, and EditDuet's failures include critic requests its tools cannot execute, such as inserting transitions \citep{sandovalcastaneda2025editduet}. These observations motivate an interface that lets the agent compose the content changes available to manual editing.

Second, the environment must provide \emph{structural validation and edit-aware evidence}. Recent systems check clips, typed timelines, and task constraints or inspect rendered previews \citep{zhao2026cutclaw,wang2026refinecut,yan2026crayotter}. In an editable project, feedback must identify violations in relationships among clips, attachments, and transitions so the agent can correct its edits. Moving or retiming material also changes where its words and shots occur, so evidence for subsequent decisions must remain aligned with the current timeline.

Third, the output must remain \emph{directly editable}: after each agent run, users must be able to revise its result and undo committed changes. The project and its edit history must therefore persist across handoffs between the user and the agent.

We introduce \textbf{Cassette-Editor}, an agent harness built on shared timeline semantics (Figure~\ref{fig:aci-overview}). Manual gestures and agent patches use one typed action language and executor, which derives timeline consequences such as shifting later clips and their attached titles after a trim. A shared \emph{rulebook} checks the resulting project and returns diagnostics that identify affected entities and fields. To ground subsequent edits in this updated state, edit-aware reads map transcripts and shot descriptions from source media to their locations in the current timeline. Committed changes enter a shared history, allowing users to revise or undo agent edits through the editor. The model chooses edits and responds to feedback, while the editor executes them, checks their validity, and maintains the project and its evidence. The harness works with released models without task-specific training.

Our contributions are summarized as follows:
\begin{itemize}
  \item \textbf{A shared human--agent editing harness}: Cassette-Editor gives manual gestures and agent patches a shared action language, executor, rulebook, and history. The model chooses edits; the editor derives their timeline consequences, checks project invariants, and maps source evidence to the current edit, while users can revise and undo committed changes.
  \item \textbf{Empirical evaluation}: component ablations and end-to-end comparisons assess completion and agent effort. With GPT-6 Luna, Cassette-Editor passes 92/93 editing trials versus 48/93 with basic tools, using one-third of the agent time and less than half the model-token cost per passed trial. On 50 TVSum videos, it achieves F1 23.8 versus 20.1 for Codex CLI using the same model.
\end{itemize}

%% file: figures/fig-aci-overview.tex
\begin{figure*}[t]
  \centering
  \includegraphics[width=\textwidth]{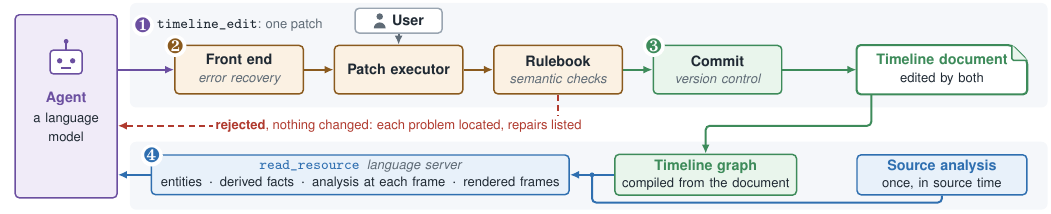}
  \caption{One write path for agent patches and manual edits; italics name the
  coding tool each stage parallels. \stepref{figagent}{1} The
  agent writes one patch. \stepref{figharn}{2} The front end reads it against
  the document schema and reports what it leaves out, the executor constructs a candidate document,
  and the rulebook checks it; manual edits enter the same executor and
  rulebook, and a blocking violation returns diagnostics without changing the
  document. \stepref{figtl}{3} Accepted retained changes commit as one history
  step. \stepref{figevid}{4} \texttt{read\_resource} answers from source
  analysis and the timeline graph compiled from the document. Version
  checks and patch translation omitted.}
  \label{fig:aci-overview}
\end{figure*}

%% file: sections/related_work.tex
\section{Related Work}

\paragraph{Editing footage with language.}
Computational editing has supported specific tasks, such as interview
cutting and dialogue-scene assembly
\citep{berthouzoz2012tools,leake2017computational}, and benchmarked shot
selection and ordering \citep{argaw2022anatomy}. Language models broaden the requests an
interface accepts: LAVE combines agent actions with a manual editor
\citep{wang2024lave}, ExpressEdit accepts language and sketches
\citep{tilekbay2024expressedit}, and EditDuet revises a draft with editing
actions and a critic \citep{sandovalcastaneda2025editduet}. Other approaches
represent timelines as identifier tokens \citep{pardo2024timelines}, source spans
\citep{sandovalcastaneda2026filmgpt}, or JSON tracks
\citep{cheng2025texttoedit}, and rendering pipelines execute plans
\citep{ding2025promptdriven,nath2026unified}, revise clips under a reviewer
\citep{zhao2026cutclaw}, or diagnose steps and redo affected segments
\citep{yan2026crayotter}. RefineCut trains a planner to revise typed timelines using verifier
feedback \citep{wang2026refinecut}. Cassette-Editor focuses on the
editing environment for an off-the-shelf model: manual gestures and
agent patches share one execution and validation path and one history
within a persistent project.

\paragraph{Action languages and feedback.}
How agents express edits and use feedback to correct them has also been
studied in software engineering. SWE-agent's linter discards edits that introduce
syntax errors, increasing the share of resolved SWE-bench Lite issues from
15.0\% to 18.0\% \citep{yang2024sweagent}.
PatchBoard represents proposed changes as JSON Patch operations, which it
checks against a schema and registered invariants before committing or
rejecting them atomically \citep{zhang2026patchboard}.
Feedback design also matters: models correct mistakes more effectively when their location is
supplied \citep{tyen2024mistakes}, although more detailed execution feedback is
not consistently better \citep{zheng2025multiturn}.
Cassette-Editor combines a shared action language for manual and agent editing
with timeline-specific execution, validation, and diagnostic feedback.
The language uses stable references because models miscompute array indices
in JSON patches \citep{duanis2025jsonwhisperer}.

%% file: sections/background.tex
\section{Background: A Timeline Project}

Figure~\ref{fig:timeline} shows the running example. A project is a timeline document: clips arranged on tracks and
measured in frames of the edit, 30 per second here. Each clip plays a window
of a source file, measured in source time, the seconds of that file: c1 plays
seconds 31.5 to 33.5 in frames 0 to 59; figures and reads write such ranges
half-open, as 0--60. A clip can be \emph{retimed}: c2 plays
at double speed, a1 plays backwards, and a2 holds one source frame, then
plays at double speed.

\input{figures/fig-timeline}

Clips relate to each other in four ways, and each makes an edit reach beyond
the clip it names; the first two follow the magnetic timeline of Final Cut
Pro \citep{apple2026connect,apple2026magnetic}. The \emph{main storyline}
on T1 plays its clips without gaps, so shortening c2 moves every later clip.
A \emph{connected clip}, such as the title on T2, is attached to a frame of
a storyline clip, here frame 15 of c1, and moves with that clip; deleting c1
deletes the title. A \emph{transition}, here a dissolve, blends two
neighboring clips across their cut and uses source footage beyond each
clip's visible trim window. A \emph{compound clip} groups clips into a
sequence of their own that the edit can use more than once: compound A is
used by two \emph{instances}, A\#1 and A\#2, so a change inside A appears
in both, and frame 21 of A plays at frames 141 and 211 of the edit.
What a viewer sees at a frame is the \emph{composite} of every clip that
plays there: frame 60, inside the dissolve, mixes c1 and c2, and the title is drawn over them
from frame 15 to 74.

%% file: figures/fig-timeline.tex
\begin{figure}[t]
  \centering
  \includegraphics[width=\columnwidth]{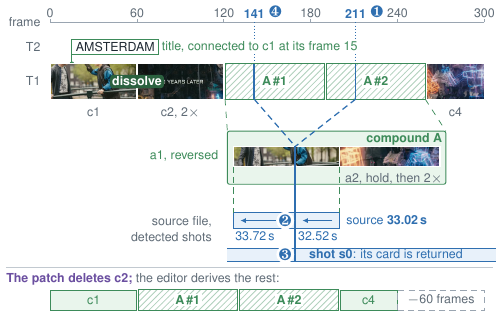}
  \caption{The running example, a 300-frame project at 30 fps.
  In blue, one read from the timeline graph (Figure~\ref{fig:timeline-graph}): \stepref{figevid}{1} frame 211 plays frame 21 of compound A,
  where a1 plays backwards, so it \stepref{figevid}{2} shows
  source second 33.02, \stepref{figevid}{3} in shot s0; \stepref{figevid}{4}
  frame 141 plays the same frame. Bottom: the storyline after a patch deletes
  c2. Frames from
  \emph{Tears of Steel} (CC BY 3.0, \copyright{} Blender Foundation,
  mango.blender.org).}
  \label{fig:timeline}
\end{figure}

%% file: sections/system.tex
\section{System Design}
Cassette-Editor combines direct timeline editing with a conversational
agent over a shared project document. The agent uses two domain tools:
\texttt{read\_resource} reads timeline and source ranges, rendered frames,
history, and writable-field declarations; \texttt{timeline\_edit} submits
patches. The model chooses edits and responds to feedback, while the
harness coordinates each \emph{run}, the agent's work on one request.
We begin by introducing the shared action language and executor, along with the rulebook and its diagnostic feedback, before detailing how edit-aware reads and shared history enable interleaved human–agent editing.

\subsection{Action Language and Execution}
\label{sec:action-lang}
We expose the document's writable state as the agent's action space,
so editing techniques can be composed from a common set of changes.
A \emph{change} creates, updates, or deletes one entity, such as a clip,
track, or transition; a \emph{patch} groups changes in one call.
Updates name fields by dotted paths, such as \texttt{source.text}, and
leave other fields unchanged. A split, for example, combines creating a
second clip with updating the clips' boundaries.

Manual gestures use the same language. A gesture's \emph{recipe}, the underlying editor code, computes canonical document changes and submits them alongside the document version as a \emph{write}; an agent patch resolves to an identical \emph{write}. The agent's language and tool declarations are
derived mechanically from the canonical document schema, and each agent
change translates to exactly one canonical change and back. The agent
can therefore name the content fields a manual write names. A new
writable content field reaches both writers without a new tool.

Reads expose this writable representation separately from derived facts.
Each entity record presents its creation fields as its \texttt{value}
and \emph{placement}. For the running example's title, these include its
text and attachment to frame 15 of c1; its frame range, 15--75, is derived.
A patch reuses the same fields to change the text or attachment
(Appendix~\ref{app:read-record}). Rejection diagnostics locate problems by those field names.

\paragraph{Interpreting patches.}
OpenAI's grammar-constrained decoding admits one JSON object in a fixed
layout while leaving its keys and values free; the edit schema applies
after decoding. The front end resolves uniquely identifiable field names
and converts compatible values. Unresolved or ambiguous input is omitted
at the smallest permissible unit, and the executor likewise omits changes
it cannot apply (Appendix~\ref{app:rulebook}). The remaining changes form the candidate
edit for checking and commit; the diagnostic receipt reports all omissions.

\paragraph{Deriving consequences.}
The document separates the relationships the model specifies from the
timing the executor derives. A clip's \emph{placement} is either a position
in the gapless main-storyline order or an attachment to a frame of a
storyline clip. The model specifies the edit without listing its dependent
timing changes. Deleting c2, for example, closes the gap, shifts both
instances of compound A and c4 sixty frames earlier, and removes the
dissolve into c2 (Figure~\ref{fig:timeline}, bottom). Manual and agent writes
follow the same execution path, so both receive these derived consequences.

Splitting a retimed clip also preserves its temporal state. Consider a
60-frame clip at 30 fps: its first half plays source seconds $[1,2)$ at
$1\times$, and its second half plays $[2,4)$ at $2\times$. A split at frame
30 creates the right part and shortens the original, using the same two
changes for the manual gesture and agent patch. The agent can specify the
retained local ranges, from which the executor derives source windows and
speed segments, rebases keyframes, and continues effect clocks across the
cut. A keyframe at local frame 12 becomes frame $-18$ in the right part,
retaining its influence on that interval. Both writes produce the same two
30-frame clips; Appendix~\ref{app:parity} traces their execution.

\subsection{Rulebook and Diagnostics}
\label{sec:rulebook}
The rulebook is the editor's executable semantic checker and linter.
It checks candidate edits against invariants from the document and temporal
model, such as acyclic nesting and source windows within their media;
editing semantics, such as one transition per cut; and the authority
model, which restricts changes to the run's edit scope and media access.

\paragraph{Checking and commit.}
Validation is part of the shared commit path
(Figure~\ref{fig:aci-overview}). After checking the document version, the
executor applies the retained changes to a copy of the document, and the
rulebook checks the result. If no blocking violation remains, the changes
commit atomically as one history step; rejection leaves the project
unchanged. Authority comes from the authenticated run, so a user can
protect a shot from the agent. Each of the rulebook's 110 rules specifies
its checking stage and outcome (Appendix~\ref{app:rulebook}): designated fixes are applied,
other errors reject, and warnings are reported. Eleven rules repair
violations caused as side effects but reject those written directly by
the patch.

Manual and agent writes use the same rule registry (Appendix~\ref{app:rulebook}).

\paragraph{Diagnostic feedback.}
A \emph{receipt} reports the outcome, corrections, and omissions from the
original input. Each diagnostic identifies the entity and field, with
typed repair options and reads that would supply missing facts. Applied
fixes are reported; repairs with multiple reasonable alternatives are
offered as options for the model to choose. After rejection, an
\emph{amend} replaces or removes selected changes, closing a
compiler-style feedback loop (Figure~\ref{fig:compiler-run}). Warnings,
such as duplicated audio, also enter the receipt. A final reply with an
open warning returns to the agent once to resolve it or explain why it
is retained; remaining warnings reach the user.
\input{figures/fig-compiler-run}

\subsection{Human--Agent Editing Workflow}
\label{sec:workflow}
The agent reads the current edit before choosing and submitting a patch.
The write path above executes and checks that patch; shared history then
makes the committed result available for subsequent manual revision.

\paragraph{Reading the current edit.}
Analysis results use timestamps in the original media and are reused
across timeline edits. At upload, a decoding pass measures shot cuts,
voice activity, sound tags, beats, and loudness. On request, Gemini 3.5
Transcribe produces word-timed transcripts, and the user's chosen model
(GPT-6 Luna in our experiments) describes content, text, and actions in
sampled frames of each shot.
The agent and analysis models run as released.

The editor compiles the project into a \emph{timeline graph}
(Figure~\ref{fig:timeline-graph}). Each sequence has its own graph, with
tracks containing ordered clips and edges for attachments, transitions,
media uses, and nested sequence instances. Compilation derives frame
ranges, mappings into nested sequences, and each instance's source path.
Reads use these results with the source-mapping, geometry, and layout code
shared with the rulebook, preview, and renderer.
\input{figures/fig-timeline-graph}

A read of frame 211
(Figure~\ref{fig:timeline}, blue) follows A\#2 into compound A, where a1
plays backwards, to source second 33.02 and returns that shot's
description. Its \emph{instance path}, \texttt{root > A\#2 > a1},
distinguishes this use from frame 141, where A\#1 plays the same source
frame. A range read returns spans with the same active layers and a record
of what each layer plays. A rendered-frame read returns the composite
image, contributing layers, and, for a marked subject, its tracked box
projected through the clip and enclosing-instance transforms. Table~\ref{tab:evidence-boundaries}
in Appendix~\ref{app:read-outputs} summarizes these read outputs; Appendix~\ref{app:read-mapping} details the temporal and spatial
mappings.

\paragraph{Taking turns and revising edits.}
Each read carries the document version, and references copied from it are
checked against both that version and what the run was shown
(Appendix~\ref{app:parity}). A run holds the write \emph{lease}, the sole right to commit,
until it ends or pauses; manual edits and undo are refused during that time
(Figure~\ref{fig:history},~\stepref{figharn}{1}). A patch not based on the
run's own reads of the current version is refused until the agent
rereads~\stepref{figharn}{2}. If the user edits during a pause, that run
cannot continue.

Accepted patches and manual edits share one history, with each accepted
write recorded as one step. The agent can undo whole steps regardless of
who made them. An edit after an undo starts a branch that preserves the
undone steps~\stepref{figtl}{3}, and the user can switch
back~\stepref{figtl}{4}. To revert one earlier step while keeping later
edits, the agent reads the document before that step and writes a forward
patch. For example, it can restore Hannah's title to its position before
a manual move by writing the earlier position into the current document.
\input{figures/fig-history}

Together, the shared action language, schema, execution path, and history
establish \emph{editing parity} for document content. Agent patches can
encode the document changes emitted by manual gestures. Under the same document version and
authority, equivalent writes share execution and validation and yield the
same content, subject to the writer-specific policies in Appendix~\ref{app:rulebook}.
Accepted edits enter the
same project document and remain directly revisable by either writer.
Appendix~\ref{app:parity} gives further editing examples and details of the history protocol.

%% file: figures/fig-compiler-run.tex
\begin{figure}[t]
  \centering
  \includegraphics[width=\columnwidth]{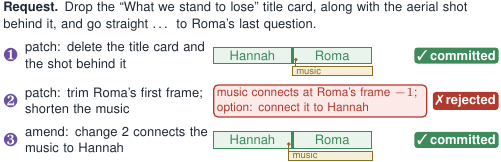}
  \caption{The full harness's only rejection in the main experiment, as
  recorded. \stepref{figagent}{1} The editor closes the gap and removes the
  dissolve into Roma's question. \stepref{figagent}{2} The trim would connect
  the music one frame before the question; the receipt names the music's
  placement and offers Hannah's answer, which plays that frame.
  \stepref{figagent}{3} The agent amends only that change.}
  \label{fig:compiler-run}
\end{figure}

%% file: figures/fig-timeline-graph.tex
\begin{figure}[t]
  \centering
  \includegraphics[width=\columnwidth]{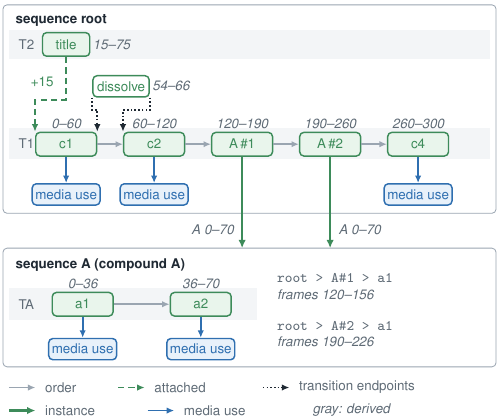}
  \caption{The timeline graph the editor compiles from the running example.
  The two instances in the root link to one graph for compound A.
  Panel and track nesting show containment; gray annotations give derived
  frame ranges, instance mappings, and source paths.}
  \label{fig:timeline-graph}
\end{figure}

%% file: figures/fig-history.tex
\begin{figure}[t]
  \centering
  \includegraphics[width=\columnwidth]{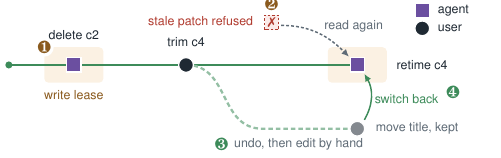}
  \caption{History of the running example; each step is one commit.
  \stepref{figharn}{1} The run deleting c2 holds the write lease.
  \stepref{figharn}{2} Retiming c4 from the version before the manual trim is
  refused, then committed after rereading. \stepref{figtl}{3} The user undoes
  the retime and moves the title, creating a branch.
  \stepref{figtl}{4} Switching back restores the retime and keeps the
  title-move branch.}
  \label{fig:history}
\end{figure}

%% file: sections/evaluation.tex
\section{Evaluation}

Internal comparisons examine capability coverage (small catalog),
combined execution support (basic tools), and component effects on
completion and effort. Injected failures test repair acceptance and cost;
TVSum compares complete systems against external human annotations.

\subsection{Experimental Setup}
\input{tables/harness}
\input{tables/components}

\paragraph{Projects and requests.}
The starting projects are two interview documentaries we assembled in the
editor from CC~BY footage: one on imaging a black hole (3.4 minutes, 37 clips
on four tracks) and a studio conversation on floods (3.2 minutes, 31 clips).
We wrote 31 requests with specified acceptance checks in five families that
probe the design decisions rather than sample typical use: \emph{find} (6 requests), placing or removing
material by what is said or shown; \emph{consequences} (9), edits that move
later clips and attachments, such as cutting a line and closing the gap;
\emph{structural validity} (6), requests operating near structural
validity boundaries, such as one-second dissolves at cuts without
enough spare footage or a title with a character its font lacks; \emph{composite} (6), J-cuts, slow motion ending on a freeze,
and picture-in-picture; and \emph{undo} (4), reverting one earlier change
while keeping later ones. They probe, in order, reads mapped to the edit,
derived consequences, the rulebook, composable changes, and the shared
history (Appendix~\ref{app:evaluation} lists every request).

\paragraph{Conditions.}
All editing conditions retain the shared project and write path, using
the same model, requests, and scorer. Five conditions reduce support:
\emph{Small catalog} replaces typed
patches with ten predefined operations (Appendix~\ref{app:evaluation}); each runs the corresponding manual action
through the same executor and rulebook. \emph{Reduced rule enforcement} keeps only the
rules a document needs to render and the authority rules, and removes the
others' checks, their statements in the prompt, and the font fixes, such as
drawing characters a font lacks, so it tests rule enforcement with these
fixes; the executor's other fixes remain. \emph{No derived consequences} leaves dependent changes, and the
editor's fixes, to the agent: where a rule needs one, the patch is rejected
with what to write, and elsewhere it commits with content misaligned. \emph{Source-time
reads} returns transcripts and shot descriptions in source time instead of
the frames of the current edit. \emph{Basic tools} combines all four and
reduces each diagnostic to its code and rule statement. Individual interventions run on targeted and control families
(Table~\ref{tab:components}).

\paragraph{Evaluation metrics.}
A trial \emph{passes} when the final document parses, its media references
resolve, it differs from the start, and it satisfies every scored request
check. This verdict measures the scored request outcomes, including preservation a
request implies near the edit, such as remaining cutaways; perceptual quality and unrequested changes elsewhere
are outside its scope. Checks read the document: the placement of a requested caption, cutaway, or inset uses temporal IoU of at least 0.5, following temporal grounding benchmarks
\citep{gao2017tall}; quantities a request states, such as one second or half speed,
are compared frame by frame; word coverage comes from the transcript;
and two models check visual labels, with ambiguous checks excluded
(Appendix~\ref{app:evaluation}). Before the runs, we validated the scorer on reference edits,
alternative edits, unchanged projects, and deliberately incorrect edits.
Raising the temporal IoU threshold to 0.7 changes no verdict across the 1{,}927 editing
trials; every passing placement has IoU at least 0.84.
We also count write calls, reads,
rejections, and the blocking violations the full rulebook, run as a check
only, finds in each final document. Agent time runs from request to final
reply, after import and source analysis and without rendering full video clips; cost prices
agent-model tokens.

\paragraph{Procedure and analysis.}
We use GPT-6 Luna as released, at high reasoning effort, for agent runs and
shot descriptions. Each trial copies the starting project and runs one turn
on the verbatim request without follow-up.
Three repetitions per request and condition yield 429 trials in randomized
blocks. Primary editing tests use task-level means and two-sided
exact sign-flip tests; Holm correction ($p_\mathrm{Holm}$) covers five editing comparisons and
the repair acceptance test (exact McNemar). Other measures are uncorrected, with \emph{exploratory} ones marked.

\subsection{Capability Coverage and Request Completion}

The full harness satisfied the request checks in 92/93 trials and basic
tools in 48/93, a difference of
47 percentage points ($p_\mathrm{Holm} = 0.0002$; Table~\ref{tab:harness}).
Basic tools used 4.5 write calls per trial against 1.3; per trial it took
1.61 times the agent time ($p = 0.0026$) at 1.15 times the cost ($p = 0.27$),
and counting failed trials, 3.01 and 2.25 times per passed trial. It alone left blocking violations:
overlapping clips in five final documents, four from failed trials.

\paragraph{Action-language ablation.}
Replacing the shared action language with a ten-operation catalog reduced
passes from 92/93 to 65/93 in a separate experiment across all 31 requests,
with the other harness components unchanged
($p = 0.0005$, its one primary test). The catalog failed every trial of eight requests, each
needing what its ten operations cannot set: a fade, a hold, inset placement,
a transition's sound, or a title's position; on the other 23 it passed 65/69
against 69/69. The catalog averaged 4.4 writes against 1.4 and took 1.36
times the agent time ($p = 0.008$) at 1.08 times the cost ($p = 0.48$).
After excluding these eight requests, the full harness passed 68/69 trials
compared with 48/69 for basic tools. On this subset, basic tools used
1.86 times the agent time and 1.40 times the model-token cost per passed
trial (Appendix~\ref{app:evaluation}).

\subsection{Component Effects on Completion and Effort}

In the main experiment, no individual intervention significantly changed its
primary measure (Table~\ref{tab:components}A).
Follow-ups used new requests at high and zero \emph{exposure}, how much
an edit depends on the removed component: a trim with clips connected
after it against the same trim with none. Each test covers all of a
component's follow-up requests, with multiplicity correction over the four
tests (Table~\ref{tab:components}B; Appendix~\ref{app:evaluation}).

\paragraph{Deriving dependent changes.}
Requests trimmed answers, with 5 or 15 clips connected after the cuts
at high exposure and none at zero. Without derived
consequences, high-exposure passes fell from 230/232 to 199/232, on 18 of
32 requests and rising on none; zero-exposure trials passed 135/136
against 136/136 ($p_\mathrm{Holm} < 0.001$ for the
difference-in-differences). All high-exposure failures left connected clips off their words.
Agent time rose 1.37-fold at high exposure and
1.05-fold at zero ($p_\mathrm{Holm} < 0.001$ for their ratio).

\paragraph{Enforcing editing rules.}
High exposure tested fonts missing requested characters and dissolves with
insufficient spare footage, which half the dissolve requests did not mention.
With reduced rule enforcement,
high-exposure requests passed 10/48 against 46/48 with the full rulebook
($p_\mathrm{Holm} < 0.001$), at 0.59 times the agent time: no title trial
passed, and dissolve trials passed 10/12 where the request named the limit
and 0/12 where it did not; there, all ten full-harness passes kept the full
second by slipping the card. Zero-exposure controls did
not decline. Failures left characters
undrawn or dissolves holding frames. Both draw only warnings,
so the main experiment's blocking-violation count stayed zero while
passes in the structural validity family fell from 94\% to 50\% ($p = 0.25$). In main title trials, a prompt stating the rule did not prevent
violations: without derived consequences, which keeps the rule and its
statement but not the fixes, all six first patches were rejected, then
passed from the diagnostic (Appendix~\ref{app:evaluation}). The effect is that of the rules' prompt statements,
checks, and font fixes together; no run inspected rendered video.

\paragraph{Mapping evidence to the edit.}
Edit-aware reads reduced agent time with little change in request-check
passing, both with and without retiming. On 23 follow-up requests, source-time reads
passed 143/144 trials against 144/144
at 1.36 times the agent time ($p_\mathrm{Holm} < 0.001$): 1.31 on
normal-speed and 1.35 on retimed answers. Main trials likewise passed 53/54
in both conditions, with 0.74 more reads without mapping ($p = 0.013$). On nested, reused clips, the full harness passed 19/24 trials against
16/24 without instance paths, which added 2.5 reads per trial (no
significant pass change; exploratory; Appendix~\ref{app:evaluation}).

\paragraph{Providing actionable diagnostics.}
The full harness was rejected once in 93 main trials
(Figure~\ref{fig:compiler-run}), so diagnostics are tested on 60 rejected
patches made by four mutations, allowing three repair attempts from the
same state. \emph{Repair acceptance} means an accepted commit within those
attempts; fidelity to the original editing request is unscored. Codes and
rule statements alone yielded acceptance in 56/60 cases against 58/60
with full receipts ($p_\mathrm{Holm} = 1.00$) and used 2.15 times
the input tokens, 2.44 times the cost, and 2.10 times the repair time
(paired geometric ratios; 54 pairs for time). With codes alone, accepted
patches equaled the unmutated one in 20 cases against 21 and added changes
outside the diagnosed locations in 12 (143 fields) against 3 (16). The observed benefit is
lower repair cost and time, with similar acceptance rates.

\subsection{Comparison with General Agents on TVSum}

We chose TVSum, an established benchmark for video summarization, because its
human importance annotations and public evaluation code enable
reproducible scoring against human selections
\citep{song2015tvsum,song2019tvsumcode}. It contains 50 videos, each with
20 annotators' scores for every two-second shot.
Given a silent video and its title, each system delivers a
summary of about 15\% of the frames in source order, cut at normal speed.
Codex was the top harness in
three of four AgenticVBench families \citep{cao2026agenticvbench},
and Claude Code with Claude Opus 5.5 ranks first on Timeline-Bench
\citep{timelinebench2026leaderboard}. We run Codex CLI
\citep{openai2026codexcli} with the same model and effort, and Claude Code
\citep{anthropic2026claudecode} at high effort with Opus 5.5 or Claude
Fable 5.2 \citep{anthropic2026fable}; all share the editing instruction, summary frame budget, and
delivery check (Appendix~\ref{app:evaluation}). Published summarizers use
other protocols \citep{barbara2025prompts,guo2026agentic}, so we compare
with random and human selections. Rendered summaries allow a common
delivery check against human references. The systems differ in interface, for Claude Code in
model, and in observations: Cassette-Editor receives shot descriptions
computed at import, while the baselines extract and view frames. The
comparison therefore measures complete-system performance.

\input{tables/tvsum}

Cassette-Editor agreed with the annotators more than Codex, F1 23.8 against
20.1, higher on 33 of 50 videos (paired permutation test,
$p_\mathrm{Holm} < 0.001$), and Claude Code with Opus 5.5 (22.0, higher on
31, $p_\mathrm{Holm} = 0.026$), but not significantly more than
with Fable 5.2 (23.1, $p_\mathrm{Holm} = 0.23$). All systems delivered valid
summaries on all 50 videos. Their F1 scores exceeded random selection
(14.7) and were below human--human agreement (25.0; Table~\ref{tab:tvsum}).
Appendix~\ref{app:evaluation} reports a supplementary importance-score analysis with the same
system ranking.

%% file: tables/harness.tex
\begin{table*}[t]
  \caption{The full harness against basic tools on 31 requests, three trials each. Passes satisfy the request checks. Per passed trial: totals over all
  trials, including failures, divided by passes. Last row: basic tools minus full harness (differences; ratios for time and cost).}
  \label{tab:harness}
  \centering
  \small
  \setlength{\tabcolsep}{4.6pt}
  \begin{tabular}{@{}lccccccc>{\columncolor{figevid!7}}c>{\columncolor{figevid!7}}c>{\columncolor{figevid!7}}c>{\columncolor{figevid!7}}c@{}}
    \toprule
    & \multicolumn{6}{c}{Pass rate by request family (\%) $\uparrow$} & & \multicolumn{2}{c}{Per passed trial $\downarrow$} & \multicolumn{2}{c}{Per trial $\downarrow$} \\
    \cmidrule(lr){2-7}\cmidrule(lr){9-10}\cmidrule(l){11-12}
    Condition & Find & Conseq. & Validity & Comp. & Undo & All & \stack{All 3 trials\\passed $\uparrow$} & \stack{Agent\\time (s)} & \stack{Cost\\(USD)} & \stack{Write\\calls} & \stack{Rejec-\\tions} \\
    \midrule
    Basic tools & \textbf{100} & 85 & 17 & 6 & 25 & 52 & 15/31 & 221.4 & 0.036 & 4.46 & 0.35 \\
    \rowcolor{figtl!10}
    \textbf{Full harness} & \textbf{100} & \textbf{100} & \textbf{94} & \textbf{100} & \textbf{100} & \textbf{99} & \textbf{30/31} & \textbf{73.6} & \textbf{0.016} & \textbf{1.31} & \textbf{0.01} \\
    \midrule
    Difference & 0 & $-$15 & $-$78 & $-$94 & $-$75 & $-$47 & $-$15 & 3.01$\times$ & 2.25$\times$ & +3.15 & +0.34 \\
    \bottomrule
  \end{tabular}
\end{table*}

%% file: tables/components.tex
\begin{table*}[t]
  \caption{Component interventions. A: primary tests on the main requests, 15 targeted and 6 control per condition (12 and 6 for source-time reads; all 31 for the small catalog in a separate experiment with one test); Holm correction includes basic tools and repair acceptance. The separate catalog experiment reports its single primary test's unadjusted $p$. B: tests on all follow-up requests, Holm-corrected together; exposure is how much an edit depends on the affected component, and zero-exposure controls make the same edit without that dependence. Arrows show full $\to$ condition; time ratios show condition/full. Pass tests compare high- and zero-exposure effects at request level; the time contrast for derived consequences compares their ratios. Source-time reads use the paired time ratio over all requests. Bold: $p<0.05$ under the stated correction.}
  \label{tab:components}
  \centering
  \footnotesize
  \setlength{\tabcolsep}{4pt}
  \begin{tabular*}{\textwidth}{@{\extracolsep{\fill}}llcc@{}}
    \toprule
    \multicolumn{4}{@{}l}{\textbf{A. Component interventions on the main requests}} \\
    Condition & Primary measure & Full $\to$ condition & $p$ \\
    \midrule
    Small catalog & Targeted pass rate (\%) $\uparrow$ & $100\to67$ & 0.31 \\
                  & Pass rate (\%), all 31 requests (separate experiment) $\uparrow$ & $99\to70$ & $\mathbf{<0.001}$ \\
    No derived consequences & Rejections/trial $\downarrow$ & $0.02\to0.17$ & 0.38 \\
    Reduced rule enforcement & Blocking violations left $\downarrow$ & $0\to0$ & 1.00 \\
    Source-time reads & Targeted pass rate (\%) $\uparrow$ & $100\to97$ & 1.00 \\
    \bottomrule
  \end{tabular*}
  \par\smallskip
  \begin{tabular*}{\textwidth}{@{\extracolsep{\fill}}lc lcc c@{}}
    \toprule
    \multicolumn{6}{@{}l}{\textbf{B. Follow-up tests of dependencies, structural validity, and read mapping}} \\
    Condition & Requests & Measure & High exposure & Zero exposure & $p_\mathrm{Holm}$ \\
    \midrule
    No derived consequences & 32 & Passes $\uparrow$ & $230/232\to199/232$ & $136/136\to135/136$ & $\mathbf{<0.001}$ \\
                            & 32 & Time ratio $\downarrow$ & $1.37\times$ & $1.05\times$ & $\mathbf{<0.001}$ \\
    Reduced rule enforcement & 16 & Passes $\uparrow$ & $46/48\to10/48$ & $47/48\to48/48$ & $\mathbf{<0.001}$ \\
    Source-time reads       & 23 & Time ratio $\downarrow$ & \multicolumn{2}{c}{All: $1.36\times$; passes $144/144\to143/144$} & $\mathbf{<0.001}$ \\
    \bottomrule
  \end{tabular*}
\end{table*}

%% file: tables/tvsum.tex
\begin{table}[t]
  \caption{Complete-system comparison on TVSum, 50 videos, one run each.
  F1 is averaged over the 20 annotators' reference summaries,
  each using a 15\% frame budget. Human--human agreement is the mean
  pairwise F1 between different annotators' reference summaries.}
  \label{tab:tvsum}
  \centering
  \small
  \setlength{\tabcolsep}{5pt}
  \begin{tabularx}{\columnwidth}{Xr}
    \toprule
    Selection & F1 (\%) $\uparrow$ \\
    \midrule
    Random 60-frame units & 14.7 \\
    Evenly spaced units & 15.2 \\
    \midrule
    Codex CLI, GPT-6 Luna & 20.1 \\
    Claude Code, Claude Opus 5.5 & 22.0 \\
    Claude Code, Claude Fable 5.2 & 23.1 \\
    \rowcolor{figtl!10}
    \textbf{Cassette-Editor}, GPT-6 Luna & \textbf{23.8} \\
    \midrule
    Human--human agreement & 25.0 \\
    \bottomrule
  \end{tabularx}
\end{table}

%% file: sections/limitations.tex
\section{Discussion}

\paragraph{Dividing editing responsibilities.}
The component experiments connect the division of responsibility to
concrete outcomes: the executor and rulebook improve request completion,
while edit-aware reads reduce agent time and diagnostic receipts reduce
repair cost. The editor maintains the relationships and observations
defined by its document semantics; the model interprets requests and
chooses edits.

\paragraph{Autonomy and expressiveness.}
Each main trial was one unattended turn revising a finished edit.
Every run reads the current version before writing (Appendix~\ref{app:parity}).
Error accumulation over long sustained turn-taking and its benefits
to editors remain unmeasured, as does full-piece assembly beyond TVSum's silent frame selection. Editing
parity covers this editor's document content (Appendix~\ref{app:parity}); judgment,
speed, skill, and other editors' features lie outside its scope.

%% file: sections/conclusion.tex
\section{Conclusion}
Cassette-Editor shows how a released language model can edit the same
persistent project as its user. The evaluation finds higher request
completion and lower agent effort, alongside stronger agreement with human
annotations on TVSum. These results support placing operational editing
knowledge in the environment while leaving edit selection to the model.
Shared project state and history keep the agent's work available for
inspection, revision, and undo.

%% file: sections/appendix.tex
\appendixsection{Read Outputs}{app:read-outputs}

Table~\ref{tab:evidence-boundaries} summarizes the outputs of the three
read types used in the human--agent editing workflow.

\input{tables/evidence-boundaries}

\appendixsection{Evaluation Details}{app:evaluation}

\paragraph{Sources.}
The black-hole film uses \emph{Event Horizon Telescope Collaboration,
B-roll: Interviews (part 1)} and \emph{B-roll: Scientists and Technology},
European Southern Observatory, CC BY 3.0; \emph{ALMA aerial view}, M.~Struik
(CERN)/ALMA (ESO/NAOJ/NRAO), CC BY 4.0; and \emph{Wholesome}, Kevin MacLeod,
CC BY 4.0. The flood film uses \emph{Interview on extreme weather with
physical geographer Hannah Cloke}, The Royal Society, CC BY 3.0; \emph{Aerial
video of the aftermath of the flood in Aqqala}, Mohammad Moheimany, CC BY
4.0; two clips of tidal flooding at the Tidal Basin, Washington DC, G.~Edward
Johnson, CC BY 4.0; \emph{Downburst Piracicaba 2025}, A.O Mapping, CC BY 4.0;
\emph{Storm clouds, Chicago}, Julian, CC BY 3.0; and \emph{Windswept}, Kevin
MacLeod, CC BY 3.0. Each source was imported and analyzed once, with shot descriptions at high reasoning effort; every trial
starts from a copy of the project whose timeline, transcripts, and shot
descriptions are checked against the template. Before an undo request, two
manual edits are committed to the copy.

\begin{figure*}[t]
  \centering
  \includegraphics[width=\textwidth]{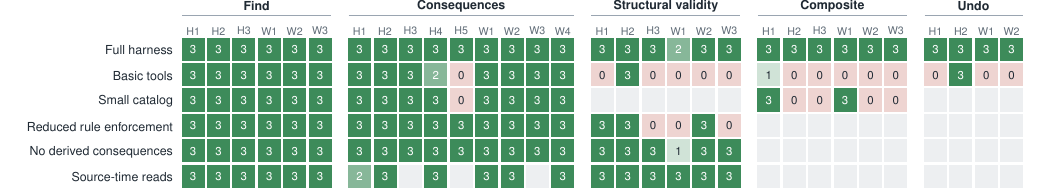}
  \caption{Trials passed out of three for every request and condition; grey
  cells are requests a condition does not run.}
  \label{fig:tasks}
\end{figure*}

\paragraph{Requests.}
H marks the black-hole film and W the flood film, numbered within each
family as in Figure~\ref{fig:tasks}; requests are quoted as sent.
Figure~\ref{fig:families} summarizes results by family.

\input{figures/fig-families}

\paragraph{Find.}
H1~When the Science Council chair says ``a telescope the size of the Earth'', put up a caption that reads ``An Earth-sized telescope''. Bring it in as he starts that phrase and keep it on screen until he finishes the sentence.
H2~When the council chair says ``And then we look at our first source and we see that ring'', cut away to the shot of an audience looking at the black-hole image on a big screen, for the length of that sentence.
H3~Take out every cutaway over the interviews that shows more than one person. Leave the other cutaways in.
W1~When Hannah first says ``flash floods'', put a caption on screen that reads ``Flash floods''. It should come up as she starts saying those words and go away when she finishes that sentence.
W2~Take out the behind-the-scenes shots where you can see the film crew and their cameras. Keep everything else as it is.
W3~In the slow-motion storm shot, a pale sheet of something blows into the top right corner of the frame. Put a caption reading ``Downburst'' on screen from the moment it first appears until the storm shot ends.

\paragraph{Consequences.}
H1~In the council chair's first answer, cut the line ``We've seen how it actually seems to do things.'' and close up the gap.
H2~In Part 1, swap the first two answers so that the astronomer speaks before the director.
H3~In the astronomer's second answer there is a run of three cutaways back to back. Drop the first one, the man at the control desk, and let the other two move up so the run still starts where it did.
H4~Shorten the director's first answer so that it ends after ``\ldots{}the most mysterious object in the universe.'' and close up the gap.
H5~Keep the music only under the opening aerial shot. It should fade out by the time the Part 1 card comes up.
W1~Cut Roma's first question, the one about water being a devastating force. Hannah's answer should follow straight after the ``When water turns dangerous'' card. Keep everything else as it is.
W2~In the ``Forecasting floods'' section, swap the two exchanges: Roma's question about what time scales they can predict on, with Hannah's answer, should come first, then the question about how good we are at predicting the next flood, with its answer. The B-roll should stay with the words it covers now.
W3~Drop the ``What we stand to lose'' title card, along with the aerial shot behind it, and go straight from the end of the flash-floods answer to Roma's last question.
W4~Trim the start of Hannah's answer about forecasting models so it begins at ``so we have huge computer models'', dropping ``We've got some really good models now''. The clouds shot in that answer should stay on the same words.

\paragraph{Structural validity.}
H1~Add a one-second cross-dissolve at every cut where one interview answer goes straight into another. If a shot doesn't have enough extra footage for a full second, make that dissolve as long as the footage allows, without freezing any frames.
H2~The music keeps playing over black after the last aerial shot ends. Make the last shot stay on screen until the music finishes.
H3~Change the opening title to: Imaging at $\lambda$ = 1.3 mm
W1~Use a one-second cross-dissolve at every cut into or out of the non-interview shots on the main picture: the opening aerial, the three section title cards and the closing clouds.
W2~Let the closing clouds shot run five seconds longer at the end, with the closing title staying up and the music carrying on until the picture ends.
W3~Change the opening title to: CO$_2$ AND WEATHER

\paragraph{Composite.}
H1~Make a J-cut into Part 2: let the council chair's voice start about a second before we cut from the Part 2 title card to him.
H2~Slow the cutaway of the people working on the big metal cylinder, in the director's answer in Part 2, to half speed, and end it on a one-second freeze before it cuts back to the director.
H3~When the director says ``We've made the first picture of a black hole'', show the shot of the black-hole image being presented on the big screen as a small picture-in-picture in the top-right corner for about three seconds, fading it in and out.
W1~Make a J-cut into Roma's next line after the ``Forecasting floods'' card: her voice should come in about a second before we cut to her, while the flooded trees are still on screen.
W2~Play the aerial shot of the flooded town, the first of the three flood shots during the flash-floods answer, at half speed, and end it on a one-second freeze frame. The other two shots should follow straight after it.
W3~When Hannah says ``how much rain is falling'', show the heavy-rain storm footage as a small picture-in-picture in the top right corner for about four seconds, fading in and out, while she stays on screen.

\paragraph{Undo.}
H1~Undo my earlier change to Part 2, but keep the two captions I added in Part 3 afterwards.
H2~Put the astronomer's name title back where it was before I moved it.
W1~Put Hannah's name title back where it was before I moved it by hand.
W2~Undo the change I made to the third section's title card, but keep the caption I added after that.

\paragraph{Conditions and runs.}
The small catalog's operations are \texttt{add\_clip}, \texttt{remove\_clip}
(a storyline clip ripples later clips, a connected clip is lifted),
\texttt{move\_clip}, \texttt{trim\_clip}, \texttt{split\_clip},
\texttt{add\_text} (styled by a catalog preset), \texttt{add\_transition},
\texttt{set\_speed} (one uniform speed), \texttt{set\_volume} (one gain in
decibels), and \texttt{set\_sequence\_format}; each call is its own history
step. Reduced rule enforcement and no derived consequences target the consequence and
structural validity families, with find as control; the small catalog targets
consequences and composite edits, with find as control; source-time reads
target find and the six consequence requests that locate a phrase (H1, H2,
H4, W1, W2, W4), with structural validity as control. The conditions ran 93, 93,
63, 63, 63, and 54 trials (full harness, basic tools, small catalog, reduced
rule enforcement, no derived consequences, source-time reads), in three repetitions
of six blocks, one block per condition, with block and request order
shuffled by a fixed seed. The catalog's limits can apply to a request of any
family, and the main experiment ran it on 21 of the 31; a separate experiment
therefore ran the full harness and the small catalog on all 31 requests, 93
trials each, in three rounds of two adjacent blocks whose order alternated,
with request order shuffled by a fixed seed. Its one primary test is the
request-level difference in pass rate. Every run used GPT-6 Luna at high effort in the standard service tier, so costs use the standard prices of 29 October 2026 \citep{openai2026pricing}, in dollars per million tokens: 0.10 for input, 0.01 for cached input, 0.125 for cache writes, and
0.50 for output; a request above 272K input tokens pays 0.20, 0.02, 0.25,
and 0.75.

\paragraph{Resource summaries.}
\input{tables/component-work}
For each condition, time and token cost per passed trial equal the sum over
all trials divided by the number that pass. Their ratios compare these
aggregate quantities; they are not estimates of retry-until-success cost.
Per-trial ratios instead take the geometric mean across tasks of condition/full ratios of within-task trial means; per trial, basic tools took 1.61 times the agent time ($p = 0.0026$) and 1.15 times the cost ($p = 0.27$). Token prices
apply to agent-model calls only. Project preparation and source analysis
are shared inputs prepared before timing, and no rendering of full video clips is included. Source analysis ran once, at import: GPT-6 Luna described the 36 minutes of source video behind the two films in 18 calls, with 1.07 million input and 45 thousand output tokens, \$0.13 at the prices above; transcripts come from a separate transcription model and are not priced here.
Basic tools made 10.9 model calls per trial against 6.2 for the full harness. Its five final documents with overlapping clips came from four failed trials and one that passed. Table~\ref{tab:component-work} summarizes agent work per condition. On targeted requests of the main experiment, the small catalog used 9.9 writes per trial
against 1.0 on the five requests needing a fade, hold, or inset, and 3.3
against 1.7 on the rest, at about equal time and cost. Without derived
consequences, rejections rose from 1 in 63 main trials to 11, six for title
characters, and cost per trial rose 1.21-fold ($p = 0.019$),
while 97\% passed against 98\%. Source-time reads took 1.26 times the agent time per trial
(exploratory).

\paragraph{Rulebook example and main failures.}
For the request to change the opening title to ``Imaging at $\lambda$ =
1.3 mm'' in a font without $\lambda$, the full executor drew the character
in the nearest catalog font that has it. Without derived consequences,
the prompt states that such a patch is rejected, yet on both title requests
the first patch of every trial left the character in the title font; each
was rejected with the covering font and character range, and the agent
wrote the range itself, passing all six trials. With reduced rule enforcement, every title trial committed
a character its font cannot draw. Across main structural validity requests,
reducing rule enforcement lowered passes from 94\% to 50\% ($p = 0.25$), with no change in write calls, time, or cost. Every
flood-dissolve trial held frames; the black-hole dissolve request, which
explicitly asks for no frozen frames, passed every trial.

\paragraph{Final documents in failed trials.}
Across the main experimental conditions, 61 failed trials committed an edit.
Of these, 57 left documents accepted by the full rulebook; the other four,
under basic tools, contained overlapping clips. The full harness's only
failed main trial kept a 12-frame hold to preserve the requested one-second
length, despite the receipt's warning and repair options.

\paragraph{Requests beyond the catalog.}
In the all-request experiment, the small catalog failed all three trials of
eight requests: the music fade (consequences H5), both freezes (composite H2
and W2), both picture-in-picture requests (composite H3 and W3), the
dissolves between answers (structural validity H1) and an undo that restores a
dissolve (undo H1), whose sound its transition operation cannot set, and an
undo of a title's position (undo W1). The full harness passed 23 of these 24
trials; in the other, its inset showed the shot next to the requested one. On the other 23 requests the small catalog passed 65/69 against
69/69, with one failure each on composite H1, consequences W1, and structural validity W1 and W2, at 3.3 writes per trial against 1.5. Over all requests it made 4.4 writes per trial against 1.4
($p < 0.0001$), 4.3 reads against 4.2 ($p = 0.72$), and 0.12 rejections
against none ($p = 0.25$), and took 1.36 times the agent time per trial
($p = 0.008$) at 1.08 times the cost ($p = 0.48$); per passed trial, 1.90
times the time and 1.59 times the cost.
A title's role label, which the catalog cannot set and no frame shows, is
not scored in this experiment; scoring it, as the main experiment does, also
fails the three trials of undo W2, for 62/93 ($p = 0.0002$). To separate
capability coverage from execution support in the main experiment, a
descriptive analysis excludes the eight requests: basic tools passed none
of their 24 trials and 48/69 on the other 23 requests, against 24/24 and
68/69 for the full harness (1.86 times the time and 1.40 times the cost per
passed trial on the 23); excluding undo W2 as well leaves 48/66 against
65/66. It is not a capability-matched comparison or a hypothesis test.

\paragraph{Follow-up requests.}
Starting states are duplicates of the two films with manual edits applied, one
template per component, film, and level; the transcripts and media analysis
match the base films. A cell's measure is passing with no blocking violation in the final
document. For derived consequences and the rulebook, every request has a
high- and a zero-exposure version, and its effect is removal minus full at
high exposure, less the same difference at zero exposure. Four tests are
Holm-corrected together, each a two-sided exact sign-flip test over all of a
component's requests: this difference-in-differences in passing for derived
consequences and for the rulebook; the same contrast in log mean agent time
for derived consequences; and, for source time, the log ratio of mean agent
time over all of a request's trials. Pilots, excluded from the analysis,
checked request wording and scoring with the full harness and set the
repetitions.

Derived consequences, 32 requests: trim the head of an answer so it starts
at a quoted sentence. Eight requests trim one answer, with five captions or
cutaways connected after the cut at high exposure, in three repetitions;
eight trim three of a film's four answers, with 15 connected clips, in six;
and 16 trim three answers, with 15 connected clips, in ten repetitions at
high exposure and four at zero. Without derived consequences, high-exposure
trials passed 17/24, 39/48, and 143/160 in these three sets, 199/232 in all,
against 230/232 with the full harness, whose two failures were in the third
set; zero-exposure trials passed 135/136 against 136/136. The
difference-in-differences is $-0.16$ (18 requests negative, one positive, 13
zero), $p_\mathrm{Holm} = 0.0002$. All 33 high-exposure failures without
derived consequences left connected clips off their words, as did the full
harness's two; the zero-exposure failure cut an answer at the wrong words.
The agent took 1.37 times the agent time at high exposure and 1.05 at zero, a
ratio of 1.31 (above one on 27 of 32 requests, $p_\mathrm{Holm} < 0.0001$),
with 161 rejections against none and 1.27 times the cost at high exposure
(uncorrected).

Rulebook, 16 requests in three repetitions: eight change a title to given
text with a character such as $\lambda$, $\rightarrow$, $_2$, or $\Delta$,
and eight add a one-second dissolve at a cut; high exposure means a title
font that lacks that character, or a card with four spare frames, which
allows eight or nine frames. Four dissolve requests ask for a dissolve as
long as the footage allows. The other four, picture-only, do not mention
short footage; for them the scorer accepts either shortening the dissolve
to fit the available footage or slipping the card to supply the requested
duration, prioritizing avoidance of frozen frames. This acceptance policy
allows a duration change not explicitly offered in the request; no passing
trial used it, since all ten full-harness passes on these requests kept 25
frames by slipping the card's footage eight frames. With reduced
rule enforcement, 10 of 48 high-exposure trials passed against 46, and 48 of
48 zero-exposure trials against 47: difference-in-differences $-0.77$ (13
requests negative, three zero), $p_\mathrm{Holm} = 0.0002$. All 24 title
trials failed, as did 2 of 12 dissolve trials whose request named the limit
and all 12 whose request did not, none with a rejection; the full harness
failed two high-exposure dissolve trials and one at zero exposure, on
requests that did not name the limit. With reduced rule enforcement the
agent finished high-exposure trials in 0.59 times the agent time
(uncorrected $p = 0.015$).

Source time, 23 requests: captions, and one cutaway, placed by quoted speech.
Eight requests ran on an answer at 1.2 to 1.3 times speed and on the same
answer at normal speed, at sentences that a placement ignoring the speed
would still put inside the same answer, in three repetitions each, one of
them in six more on its retimed answer; 15 others ran on one answer, seven
retimed and eight at normal speed, in six repetitions. Source-time reads
passed 143 of 144 trials against 144 and took 1.36 times the agent time,
above one on 22 of 23 requests ($p_\mathrm{Holm} < 0.0001$): 1.35 on retimed
answers and 1.31 on normal-speed ones, where a request run at both speeds
counts once in the overall ratio, with 1.25 more reads per trial and 1.15
times the cost (uncorrected $p \le 0.001$).

\paragraph{Exploratory nesting experiment.}
This experiment tests reads through nested and reused clips and is analyzed
apart from the ablation claims. At high exposure one
answer is a compound holding a second compound, used four times: in the body, a
teaser, a recap at 1.25 times speed, and a silent slow shot; at zero exposure the
same film is plain clips. Eight requests, four per film, locate a sentence, change
one use, remove a sentence wherever heard, or title the speaker where visible. The
removal stops reads at instance boundaries; writes, receipts, and the view attached
at turn start are unchanged, so instance details can still reach the agent. A scorer
independent of the editor expands the nesting. It ran three repetitions,
96 trials. The full harness passed 19 of 24 nested trials and 23 of 24 flat ones;
without instance paths, 16 of 24 and 24 of 24. The difference-in-differences
is $-0.17$ in pass rate (four requests negative, one positive, three zero) and
a ratio of 1.09 in time, both $p_\mathrm{Holm} = 0.55$. At high exposure the agent
read 2.5 more times and made 2.7 more model calls per trial without the paths,
against $-0.3$ and $-0.2$ at zero (difference-in-differences, uncorrected
$p = 0.016$ for each). These tests do not
establish that instance paths are needed or unneeded.

\paragraph{Controlled repair acceptance.}
Two of the six mutations, running a clip into its
neighbor and putting characters outside the font into a title, no longer
produce a rejection, since the executor moves the clip and draws the
characters. The other four move a source window past the end of its media,
lengthen a transition past the footage, place a clip on a track of the wrong
kind, or move a connected-only clip into the storyline. A mutated patch becomes a case
only if the editor rejects it with a code of its class; cases are drawn
across classes by a fixed seed. Repair acceptance means that the editor
accepts a patch within three write attempts; this does not require restoring the
original patch or satisfying the original editing request. Differences
outside diagnostic locations compare the accepted patch's authored fields
with the rejected input, rather than judging their effect on the film: 14 and
19 cases differed there with the full and the reduced receipt, 3 and 12 by
adding changes (16 and 143 fields). The accepted patch equaled the
unmutated one in 21 and 20 cases.
Input tokens and token cost include all 60 paired cases. Repair time runs
from the replayed rejection to the accepted commit and includes only the
54 cases accepted in both branches. Ratios are geometric means of the
per-case ratios. These resource ratios are not Holm-corrected.

\paragraph{Reading misauthored patches.}
In 25 of the 429 main trials, the front end left out something the agent
wrote with no reading, such as a field its target lacks; 24 passed, and the
other failed on frozen frames.

\paragraph{Fault injection.}
The 24 tasks come from the editor's acceptance suite for editing
operations, run once each. In each run the first patch the agent wrote was
rewritten by one of the ways models were recorded miswriting patches, taken
in a fixed rotation and skipped when it did not apply: an abbreviated key
(4), a skipped path segment (3), a misspelled key (2), an unknown sibling key
(2), a near value (2), a name-value array (2), and one each of a field at the
wrong level, a wrong primitive, an extra nesting, an extra key, a case change,
a separator change, and an out-of-range value; two runs received none. A
task passes when the editor's own check of its document passes. A harness
defect is an input the front end fails to lower, a receipt diagnostic of the
input or execution stage, or a grammar or transport error; none occurred. The
front end left out 27 places and read 6 under another name; 5 patches wrote
nothing, and no patch was rejected. Of the 22 rewritten first patches, 14 had
something left out, 6 were read under another name, and 2 were taken as
written: an extra nesting, which names the same field, and the out-of-range
value below. After an omission the agent wrote again in 12 of the 14 runs; in
the other two, what was left out was an unknown key and a gap clip's source
equal to its default, so the committed patch already did the task. Of the two failed tasks, one check
compared colors by spelling, and in the other an injected keyframe $10^{12}$
frames away committed without the warning the rulebook now gives.

\paragraph{Scoring.}
A word counts as heard when it plays in full and as not heard when none of
it plays; a word the cut divides satisfies either reading. This boundary
convention does not verify acoustic completeness. A reference
interval anchored on words runs from the first word's start to the last
word's end, and where the two transcripts of a source place a sentence
boundary differently, either is accepted. Nine requests depend on what shots show, so their checks use content labels naming the
shots or source intervals that show the content. Before the runs, Claude Opus 5.5 and
GPT-6.1 Sol, both at high effort, each checked every label independently, describing
still frames before answering yes or no; the scorer itself calls no model. The labels
keep what both models agreed on; where they disagreed, the shot or interval counts as
ambiguous, and a check that depends on it is excluded from that trial's pass verdict rather than counted as a pass; no check was excluded in any trial. One label was kept although both models rejected its
interval as worded: its check uses only the interval's start, where the models
differed by one frame, negligible under the caption's span IoU.
Acceptance is task-specific: for example, the black-hole dissolve request
allows duration capped by available footage, whereas the flood request in
the main experiment requires one second and permits trims or slips to make
room. A timeout or a missing commit counts as a failed trial.

\input{figures/fig-read-path-detail}

\paragraph{TVSum comparison.}
All systems received the original videos with audio removed; two videos
decode one frame more than their annotations, and all received the first
$N$ frames, losslessly re-encoded at their own frame rates. Claude Code received Codex's prompt
unchanged. Cassette-Editor's requests come mechanically from it: they keep its editing instruction, normal-speed cuts with no text,
effects, transitions, holds, other media, or audio, no repeated or
overlapping source frames, and the frame budget in source frames and
seconds, and drop what concerns only Codex's directory, such as file names,
tools, and encoding settings; Cassette-Editor's request says the timeline is
the delivery. Codex and Cassette-Editor ran GPT-6 Luna at high effort, and
Claude Code ran Claude Opus 5.5 or Claude Fable 5.2 at high effort. Codex CLI 0.166.0 and Claude
Code 2.1.322 each worked in a private directory holding only the video,
without network tools, cut and rendered with FFmpeg and Python, and viewed
frames it extracted in every run. For Cassette-Editor, each video was imported and analyzed
once, and each request ran one turn on a copy of that project without
follow-up, after which the editor's local renderer exported the final
document on a Mac. A delivery passes when it is H.264 with 4:2:0 chroma at
the sequence frame rate (within 0.01), has as many frames as its cuts, shows
at every frame the source frame its cuts name (mean absolute error of
$64\times36$ grayscale frames at most 8 on average and 20 at the 99th
percentile), carries no audible track (peak at most $-60$ dBFS), and has
nonempty, disjoint, chronological source intervals totaling 14 to 15\% of
the source frames; resolution may differ from the source. A baseline's cuts
are the source-frame intervals it must also save in a file, and
Cassette-Editor's are those of its timeline document. We score with the official code, corrected in 2019 so that random summaries
score near the budget \citep{song2019tvsumcode,otani2019rethinking}. Each annotator's reference keeps the
highest-rated 60-frame units of source frames, as the official code defines
them, up to 15\% of the frames by its knapsack \citep{song2019tvsumcode}; a delivery scores the F1 overlap of its frames with each
reference, averaged over annotators, and zero if it fails the check. Random
controls take 60-frame units in random order until the budget is filled
(mean of 100 draws); evenly spaced units cover the video at equal intervals.
The human--human agreement reference averages the F1 between each
annotator's reference summary and the other 19, then averages over the
20 annotators and 50 videos.
Wins, ties, and losses compare each video's unrounded F1 between
Cassette-Editor and each baseline; the two-sided paired permutation test, as
in the ablations, flips the signs of the 50 per-video differences in two
million random draws, with Holm correction over the three baselines. A sign
test on the wins alone gives $p = 0.033$ against Codex, $0.12$ against Claude
Code with Opus 5.5, and $0.20$ against it with Fable 5.2, where Cassette-Editor won 30 of 50.
\paragraph{Supplementary importance-score analysis.}
Table~\ref{tab:tvsum-importance} reports an exploratory measure that compares
a delivery with the annotators' importance scores directly, as proposed for
estimated importance scores after segmentation was found to dominate the
standard measure \citep{otani2019rethinking}; the statistic itself is ours.
It averages the annotators' mean scores over the selected frames,
rescaled so that the video's average scores 0 and its 15\% highest-rated
frames 100; a failed delivery scores 0, the controls score the same
selections as their F1, and one annotator's knapsack summary is scored and
rescaled against the other 19's mean, then averaged over all annotators.
The system ranking matches F1. On this measure, Cassette-Editor scored higher than Codex on 32 of 50 videos (paired
permutation test, uncorrected, $p < 0.001$; sign test, $p = 0.065$), than Claude Code with
Opus 5.5 on 32 ($p = 0.011$; sign test, $p = 0.065$), and than it with Fable 5.2
on 33 ($p = 0.31$; sign test, $p = 0.033$).
\input{tables/tvsum-importance}

\appendixsection{Temporal and Spatial Read Mapping}{app:read-mapping}

Figure~\ref{fig:read-path-detail} details the read from the timeline graph in
Figure~\ref{fig:timeline-graph}, using the running example in
Figure~\ref{fig:timeline}. Played backwards, a1's local frame $k$ covers source times
$[33.72-(k+1)/30,\,33.72-k/30)$ and is sampled just below the upper end, so
frame 21 reads 33.02\,s. Besides the
source time, a rendered-frame read carries a tracked subject's box from the
source picture into the pixels of the final composition, through the clip's
fit, its transform, and the transform of each enclosing instance, after
tracking it through its shot from the frame where the agent marked it. When
a projection cannot be computed, the read says so.

\appendixsection{A Read Record and a Patch}{app:read-record}

Figure~\ref{fig:record} shows the read record of the running example's title
and a patch written in the same form. The patch copies the field names the
read prints: it changes the placement's offset and \texttt{source.text}, and
it leaves the derived frame range to the editor. In a connected placement,
\texttt{offsetFrames} counts the clip's first frame from the anchor's first
frame, and \texttt{pointFrames} is the clip's own frame at the connection
point. Trimming the anchor's head lowers the offset so the clip stays over
the same picture; if the connection point then falls before the anchor, the
clip is removed with a report, or the write is rejected when the patch wrote
that clip (Figure~\ref{fig:compiler-run}).
\input{figures/fig-record}

\input{sections/parity-details}

\appendixsection{Rulebook Details}{app:rulebook}

\paragraph{Rule inventory.}
The 110 rules form one registry, each a predicate over the candidate state,
the proposed change, and any relevant authority, with an identifier, a
diagnostic code, a severity, the stage that judges it, and its outcome.
Eighty-seven are errors and 23 warnings. They are judged in order: the change as
the executor reads it (4), the records the write resolves to (16), what the write
itself wrote (9), the settled candidate document (67), the graph compiled from
it (4), and facts only the commit service holds, such as fonts (10). Of the
errors, 65 reject: 42 because no writing of the edit can hold, 15 because more
than one repair is reasonable, 7 for authority, and one because only a stored
document can break it. Ten are always repaired, one of them only on agent writes;
11 are repaired when the patch only disturbed what they protect and rejected
when it wrote that itself; and one differs by writer. The always-repaired
rule limited to agents draws a title character in a font that has it; the
rule that differs by writer, a warning for a user and a rejection for an
agent, covers animated text its font cannot draw when a catalog font can;
and a motion design's render check, a rejecting rule, runs before an agent's
commit and after a user's. A user's motion design that fails that check stays in
the document, marked as failed: export refuses it, reads report it, and an
agent write that changes it must render. Warnings report conditions that the editor permits but
makes explicit, such as a clip that never reaches the canvas, audio that plays
twice, or black picture after the main storyline ends. An applied fix is reported in the receipt: a clip that lands on another
off the main storyline, for instance, moves with its connected clips to the
nearest free track of its side. A repair option is
offered when it clears its own problem and adds none, marked partial while
others remain. Other guarantees come from the representation: the main storyline stores
order and durations, so its clips are gapless by construction.

\paragraph{Fixes and written violations.}
Fixes run before the settled-document rules, in a fixed order, and repeat
until nothing changes or a document recurs;
the fixed document is rebuilt and judged once by every later stage, so a fix
that leaves or causes a violation rejects the write. A violation counts as
written by the patch when the patch's own changes, before any fix, created the
entity or changed the field the rule protects; a field resent with its old value
does not count. The font steps judge only text clips or fonts the write
changed, so text a user wrote earlier does not block an agent's edit
elsewhere; an agent's edit to such a clip is judged on its whole text. Reduced
rule enforcement keeps 74 of the 110 rules, 72 errors and 2 warnings.

\paragraph{Reading a patch.}
Every tool call is decoded under one context-free grammar, supplied through
OpenAI's API, that constrains only syntax: one JSON object, one field or
element per line, nested at most eight levels deep. It leaves keys,
discriminators, references, and values free, because a constraint narrower
than what the model means would steer it to a different valid input that no
receipt could report. The schema and the front end judge these after
decoding. Each change first takes one
flat form, in which a dotted key and a nested object name the same field. What
has no reading is left out at the smallest place that may be omitted: an
optional field, a list or map written whole, since writing part of one would
remove the rest, then a change and any change naming it. A change the executor
cannot apply to the document is left out the same way, and an input with nothing
left changes nothing. The changes that remain still commit, and the receipt
lists each omission, so a split missing its new clip commits as a trim; a
clip written afterwards is a plain copy, without the split's lineage or
continued effect clocks, unless the agent undoes the trim and writes the
split again. Of the 1{,}927 editing trials, 7 committed
a patch with a change left out; all seven passed, five after a further write.
A run that fails or times out before that write leaves the partial patch
committed. Each such step passed the rulebook and is one history step, so
the user receives a document the editor accepts and can undo the run's steps
one at a time or restore the document to before the run (Appendix~\ref{app:parity});
completion of the request is not guaranteed.

%% file: tables/evidence-boundaries.tex
\par\medskip\noindent\begin{minipage}{\columnwidth}
  \captionof{table}{Information returned by source, timeline, and rendered-frame reads.}
  \label{tab:evidence-boundaries}
  \centering
  \small
  \begin{tabular}{@{}p{0.27\columnwidth}p{0.67\columnwidth}@{}}
    \toprule
    Read output & Returned information \\
    \midrule
    Source analysis &
    Transcripts, shot descriptions, and audio measurements with timestamps
    and locations in the original media. \\
    Derived timeline &
    Clip placement, active layers, instance paths, source-time mappings,
    and geometry in the current edit. \\
    Rendered frames &
    Sampled composite images, contributing layers, and tracked subject boxes
    projected into composition pixels. \\
    \bottomrule
  \end{tabular}
\end{minipage}\par\medskip

%% file: figures/fig-families.tex
\begin{figure*}[t]
  \centering
  \includegraphics[width=\textwidth]{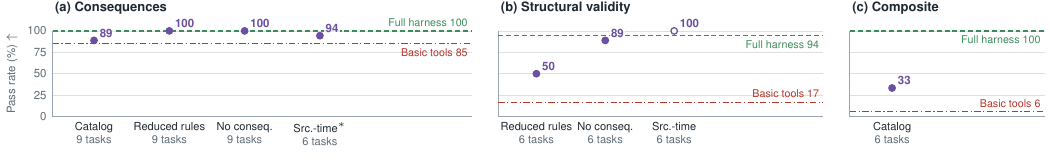}
  \caption{Pass rate by request family with component support reduced (Reduced rules: reduced rule enforcement; No conseq.: no derived consequences;
  Src.-time: source-time reads). Filled points are targeted families, hollow points control families; lines show
  the full harness and basic tools. $^{*}$The six requests that locate a phrase. Every condition passed every find trial.}
  \label{fig:families}
\end{figure*}

%% file: tables/component-work.tex
\begin{table}[!htb]
  \caption{Agent work in the component interventions on the main
  requests, against the full harness. Per-trial differences average all
  requests a condition runs, targeted and control (21; 18 for source-time
  reads); the main text's 4.4 against 1.4 writes is the all-request experiment. Time and cost per passed trial divide
  totals over all trials, including failures, by passes. Targeted pass
  rates cover the targeted requests only (15; 12 for source-time reads) and
  had no control-family decline. Not Holm-corrected;
  Table~\ref{tab:components} gives the primary tests.}
  \label{tab:component-work}
  \centering
  \footnotesize
  \setlength{\tabcolsep}{3.5pt}
  \begin{tabular}{@{}lrrrr@{}}
    \toprule
    & Small & No derived & Reduced & Source-time \\
    Measure & catalog & conseq. & rules & reads \\
    \midrule
    \multicolumn{5}{@{}l}{\emph{Per trial, difference}} \\
    \quad Write calls & $+3.10$ & $+0.06$ & $0.00$ & $0.00$ \\
    \quad Reads & $-0.22$ & $+0.05$ & $0.00$ & $+0.74$ \\
    \quad Rejections & $+0.13$ & $+0.16$ & $-0.02$ & $0.00$ \\
    \addlinespace
    \multicolumn{5}{@{}l}{\emph{Per passed trial, ratio}} \\
    \quad Agent time & $1.50\times$ & $1.07\times$ & $1.11\times$ & $1.27\times$ \\
    \quad Cost & $1.45\times$ & $1.18\times$ & $1.07\times$ & $1.14\times$ \\
    \addlinespace
    Targeted pass (pp) & $-33$ & $-2$ & $-18$ & $-3$ \\
    \bottomrule
  \end{tabular}
\end{table}

%% file: figures/fig-read-path-detail.tex
\begin{figure*}[t]
  \centering
  \includegraphics[width=\textwidth]{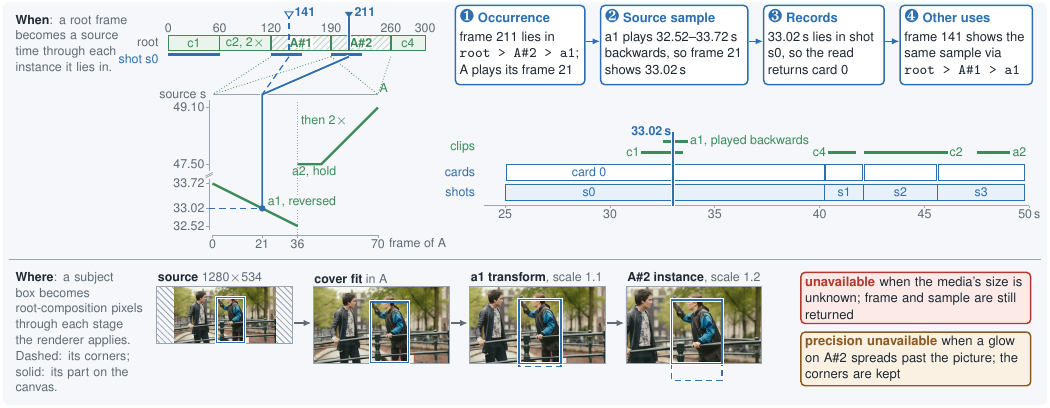}
  \caption{The read path in full, computed by the editor on the running
  example's storyline. Top: root frame 211 lies in the second instance of
  compound A, which plays its frame 21; a1 plays that frame backwards, so the
  read reports one source time, and stored evidence is found by that time;
  the steps are numbered as in Figure~\ref{fig:timeline}.
  The plot gives the source time at every frame of A. Bottom: a subject box
  is carried through cover fit, a1's clip transform, and the instance
  transform of A\#2 into root-composition pixels; when a projection cannot be
  computed exactly, the read says so. Frames, source times, and pixel
  coordinates are the editor's output; the document's windows and transforms
  are chosen, and the subject box is placed by hand.}
  \label{fig:read-path-detail}
\end{figure*}

%% file: tables/tvsum-importance.tex
\begin{table}[t]
  \caption{Supplementary TVSum importance scores (exploratory), averaged
  over 50 videos. For each video, 0 is its mean importance and 100 is the
  mean importance of its highest-rated 15\% of frames. The human reference
  scores each annotator's summary using the other 19 annotators' ratings.}
  \label{tab:tvsum-importance}
  \centering
  \small
  \setlength{\tabcolsep}{5pt}
  \begin{tabularx}{\columnwidth}{Xr}
    \toprule
    Selection & Importance $\uparrow$ \\
    \midrule
    Random 60-frame units & $-$0.1 \\
    Evenly spaced units & $-$0.8 \\
    \midrule
    Codex CLI, GPT-6 Luna & 23.3 \\
    Claude Code, Claude Opus 5.5 & 31.5 \\
    Claude Code, Claude Fable 5.2 & 36.9 \\
    \rowcolor{figtl!10}
    \textbf{Cassette-Editor}, GPT-6 Luna & \textbf{39.5} \\
    \midrule
    Human reference & 40.3 \\
    \bottomrule
  \end{tabularx}
\end{table}

%% file: figures/fig-record.tex
\begin{figure}[!t]
  \centering
  \footnotesize
\begin{verbatim}
read_resource: the title's record
{"record": "entity", "ref": "clip_8",
 "entity": "clip", "type": "text",
 "placement": {"kind": "anchored",
   "anchorClipId": "clip_5",
   "offsetFrames": 15, "pointFrames": 0,
   "track": "track_2"},
 "value": {"displayName": "title",
   "durationInFrames": 60,
   "source.kind": "text",
   "source.text": "AMSTERDAM"},
 "facts": {"sequenceId": "sequence_2",
   "timeWindow": {"startFrame": 15,
     "endFrame": 75,
     "durationInFrames": 60}}, ...}
timeline_edit: a patch of one change
{"kind": "patch", "changes": [
 {"op": "update", "entity": "clip",
  "type": "text", "target": "clip_8",
  "placement": {"offsetFrames": 30},
  "set": {"source.text":
    "AMSTERDAM, 2012"}}]}
\end{verbatim}
  \caption{A read record and a write in the same form, computed by the
  editor on the running example. The read refers to the title as
  \texttt{clip\_8}, to c1 as \texttt{clip\_5}, and to a connection as
  \texttt{anchored}. The placement and the
  \texttt{value} are what a create writes; the frame range under
  \texttt{facts} is derived. The patch moves the title's connection point
  and changes its text; the patch executor and the rulebook accept it, and
  the title now starts at frame 30. Other fields of the read are elided.}
  \label{fig:record}
\end{figure}

%% file: sections/parity-details.tex
\appendixsection{Parity on a Worked Example}{app:parity}

Table~\ref{tab:parity} shows how splitting, duplication, and audio detachment
map to shared document changes.

\input{tables/parity}

\paragraph{Splitting with temporal state.}
The project starts with a 60-frame video clip at 30 fps. Its source window
is $[1,4)$ seconds: the first 30 local frames play at $1\times$ speed, and
the next 30 at $2\times$. The clip has a position keyframe at local frame 12
and an effect stack containing an animated vignette and film grain. The user
splits it at local frame 30.

The split gesture sends a write of two changes. One creates a clip based on
the original over its local range $[30,60)$, placed right after the original
in the storyline; the other updates the original to its first 30 frames. The
gesture's recipe also states the source window, speed segments, keyframes,
and effect uses each part keeps. The agent writes the same two changes as one
patch of generic changes; it may state only the range and the new length,
from which the executor derives those fields. No split-specific tool is
involved.

Execution gives two 30-frame clips. The first plays source $[1,2)$ at
$1\times$ and retains the keyframe at local frame 12. The second plays
source $[2,4)$ at $2\times$ and retains that control point at local frame
$-18$, preserving its influence on the retained interval. Its effect uses
continue the original effect clocks across the cut. Both clips remain in the
original storyline position, with the right part after the left; from the
two changes together, the executor also derives the lineage that records the
split, and the rest of the
document is unchanged. This is the document the split's independently
written expectation specifies: losing the pre-cut keyframe, restarting the
effects, or copying the wrong source interval would leave two structurally
valid clips that do not match it.

\paragraph{Scope.}
Parity covers the document content a write changes: sequences, tracks,
clips (video, audio, image, text, graphic, compound, and multicam),
transitions, effects, color grades and LUTs, fonts, motion designs,
keyframes, speed segments, and markers. Local UI state, such as
selection and panel layout, lies outside this contract, as do identities and
facts computed by the harness.

\paragraph{Turns and history.}
A run takes the write lease when it starts and gives it up when it ends or
pauses for the user; if the user edits during a pause, the run cannot continue.
Each run starts without a write base, so a patch must follow that run's own
read of the current version, and references remembered from an earlier run are
refused until read again: the stale patch of Figure~\ref{fig:history}. Each
accepted patch is one history step and stays if the run later fails or times
out. The user undoes one step at a time or restores the document to before a
run. The agent undoes and redoes steps through a history input of
\texttt{timeline\_edit}, under the same version check and rulebook, and cannot
switch branches. References from a read of an earlier version only read; a
forward patch names current entities and writes the old values itself, so
reverting one earlier step is the model's work and must account for later
edits.

%% file: tables/parity.tex
\begin{table}[!t]
  \caption{Manual edits as the writes their recipes send, in the
  form the agent writes them, and selected facts of the document the executor
  yields.}
  \label{tab:parity}
  \centering
  \small
  \setlength{\tabcolsep}{3pt}
  \begin{tabular}{@{}p{0.19\columnwidth}p{0.34\columnwidth}p{0.40\columnwidth}@{}}
    \toprule
    \raggedright Manual edit & \raggedright Changes it sends & \raggedright Resulting document \tabularnewline
    \midrule
    \raggedright Split a retimed clip &
    \raggedright Create a copy of the second local range after the original; update the original to the first range. &
    \raggedright Source windows $[1,2)$ and $[2,4)$ s; rates $1\times$ and $2\times$; rebased keyframes and continuing effects. \tabularnewline
    \addlinespace
    \raggedright Duplicate a clip with effects &
    \raggedright Create copies of the effect entities; create a copy of the clip whose stack names the new effects. &
    \raggedright Same source, timing, and animation; distinct effect entities with matching parameters and stack order. \tabularnewline
    \addlinespace
    \raggedright Detach audio &
    \raggedright Create one audio clip based on the video; the executor derives what it copies, where it connects, and the picture's silence. &
    \raggedright Matching source window and speed segments; audio linked to the video; original video audio silenced. \tabularnewline
    \bottomrule
  \end{tabular}
\end{table}